\documentclass[pra,aps,10pt, twocolumn, floatfix]{revtex4-1} 

\usepackage[cp1251]{inputenc}
\usepackage {color}
\usepackage {graphicx} 
\usepackage {amssymb} 
\usepackage {amsmath}
\usepackage{multirow}

\usepackage{soul}

\definecolor{orange}{rgb}{1,0.7,0.4}
\definecolor{green}{rgb}{0.3,1,0.3}

\begin{document}

\title{Excitons in square quantum wells: microscopic modeling and experiment}

\author{E.~S.~Khramtsov}
\author{P.~A.~Belov}
\email{pavelbelov@gmail.com}
\author{P.~S.~Grigoryev}
\author{I.~V.~Ignatiev}
\author{S.~Yu.~Verbin}
\author{S.~A.~Eliseev}
\author{Yu.~P.~Efimov}
\author{V.~A.~Lovtcius}
\author{V.~V.~Petrov}
\author{S.~L.~Yakovlev}
\affiliation{St. Petersburg State University, Ulyanovskaya 1, 198504 St. Petersburg, Russia}

\date{\today}

\begin{abstract}
The binding energy and the corresponding wave function of excitons in GaAs-based finite square quantum wells (QWs) are calculated by the direct numerical solution of the three-dimensional Schr\"{o}dinger equation.
The precise results for the lowest exciton state are obtained by the Hamiltonian discretization using the high-order finite-difference scheme.
The microscopic calculations are compared with the results obtained by the standard variational approach.
The exciton binding energies found by two methods coincide within $0.1$~meV for the wide range of QW widths.
The radiative decay rate is calculated for QWs of various widths using the exciton wave functions obtained by direct and variational methods.
The radiative decay rates are confronted with the experimental data measured for high-quality GaAs/AlGaAs and InGaAs/GaAs QW heterostructures grown by molecular beam epitaxy.
The calculated and measured values are in good agreement, though slight differences with earlier calculations of the radiative decay rate are observed.
\end{abstract}

\maketitle
\section{Introduction}

Excitons in bulk semiconductors and heterostructures have been under intensive study for many years since their discovery in 1952~\cite{Gross}.
One of the important characteristics of an exciton is the binding energy caused by the Coulomb interaction of the electron and the hole~\cite{bib4,bib7,Ivchenko,Koch,Seisyan2012}.
Although in bulk semiconductors this energy is relatively small, typically lower than the lattice vibration energy at room temperature,
in the semiconductor heterostructures it can increase significantly, up to four times~\cite{bib4}.
Together with the binding energy, the radiative properties of an exciton are characterized by another important parameter, the radiative decay rate~\cite{Ivchenko} or the oscillator strength~\cite{Knox}, which 
is defined by the exciton-light coupling.
Since the discovery of the giant exciton oscillator strength (the Rashba effect~\cite{Rashba1962}),
the exciton states and exciton-light coupling have been drawing much attention, see, for example, Refs.~\cite{Kavokin1994,Prineas,Kazimierczuk}.
Recent developments of the microcavities~\cite{Kavokin2005,Gibbs} open up new frontiers for controlling the exciton-light coupling efficiency.

Precise experimental determination of the exciton binding energy is a quite complicated problem~\cite{Seisyan2012}.
The exciton transition can be shifted and broadened due to defects and roughness of the interfaces in a heterostructure.
The edge of optical transitions between uncoupled electron and hole is usually
inexplicit, therefore the spectroscopic determination of the exciton binding energy leads to significant uncertainties.
The most accurate methods seem to be a spectroscopic study of optical transitions to different exciton states ($1s$, $2s$, ... ).
The experimental study of exciton-light coupling and quantitative determination of its characteristics are also a nontrivial problem. The coupling gives rise to an energy shift and a radiative broadening of exciton lines~\cite{Ivchenko}, which can be, in principle, measured using, e.g., the steady-state reflectance spectroscopy, see Refs.~\cite{Poltavtsev,Poltavtsev2014} and references therein. This approach, however, is applicable only for the high-quality heterostructures when the non-radiative processes do not dominate over the radiative one. Another approach is the time-resolved spectroscopy using photoluminescence kinetics, pump-probe, or four-wave mixing methods for determination of the radiative decay time, see, e.g., Refs.~\cite{Shah,Kavokin2005,Bajoni,Portella,Singh}. Similar problems appear in this approach if the quality of the heterostructure is not high enough.

The theoretical modeling of the exciton in a bulk semiconductor is usually carried out in the framework of the hydrogen model~\cite{Knox}.
In this model, the motion of an exciton as a whole and relative motion of the electron and the hole are separated that leads to the two independent Schr\"{o}dinger equations (SEs) for the center-of-mass and intrinsic motions.
The solution of the first equation is the plane wave, whereas the latter equation is similar to the SE for a hydrogen atom and can be analytically reduced to the one-dimensional problem.
In this case, the exciton binding energy is easily calculated for various semiconductors.
For example, in the case of bulk GaAs semiconductor, this energy is $R_{x}=4.26$~meV~\cite{bib1}.

The hydrogen model becomes unsuitable for semiconductors with degenerated valence band.
In this case, the Kohn-Luttinger Hamiltonian~\cite{bibLutt}, which couples the center-of-mass and intrinsic electron-hole motions, is more appropriate.
As a result, the variables of the exciton SE cannot be separated and one has to consider the multi-dimensional SE~\cite{bib1}.
A study of the exciton in a quantum well (QW) meets further complications of the problem.
Even in the simplest case, when only the diagonal part of the Kohn-Luttinger Hamiltonian is included in the problem, the presence of the QW potential requires one to consider at least the three-dimensional SE, which cannot be solved analytically.

The exciton states in QW have been theoretically and numerically studied by many authors.
The pioneering work of Miller~\textit{et~al.}~\cite{Miller} has considered the two-dimensional exciton approximation for narrow QWs and obtained the exciton binding energy close to $4R_{x}$.
Following works of Bastard~\textit{et~al.}~\cite{bib4} and Greene~\textit{et~al.}~\cite{Greene}
employed the variational approach with various types of trial wave functions.
In these works, the QWs with finite barriers as well as with infinitely high ones were studied.
The binding energies have been calculated for different QW thicknesses and
the breakup
of the $4R_{x}$ limit has been obtained for excitons in the QWs with finite barriers.
Similar results have been obtained in the paper of Leavitt and Little~\cite{Leavitt} who have applied the method based on the adiabatic approach.
It is worth noting that these authors have also studied the indirect excitons.
More recent work of Gerlach~\textit{et~al.}~\cite{bib5} describes a phenomenological model for the binding energy of an exciton in a QW.
The authors compare the results of their model with the energies obtained by the variational approach with the prescribed trial function for the exciton wave function. 
The comparison shows a certain advantage in accuracy of the variational approach over the phenomenological model.

The trial wave functions used in the variational calculations of the exciton binding energy can be accurately chosen for two extreme cases.
One of such cases is the wide QW, when the width, $L$, is much larger than the exciton Bohr radius, $a_{B}$, so that the motion of the exciton as a whole can be separated from the intrinsic motion of the electron and the hole. 
In this case, the trial function is a product of the hydrogen-like $1s$-function and a standing wave describing the quantization of the exciton motion in the wide QW.
Another case is the narrow QW, in which the quantum confined energy of the electron and the hole is much larger than their Coulomb coupling energy.
Then, the separately obtained wave functions of the quantum confined electron and hole states are appropriate approximations~\cite{bib5}. The part of wave function taking into account the Coulomb interaction is usually modeled by a simple function with some parameters~\cite{bib5, Ivchenko}. 
So far, the precision of such approximations has not been studied in detail.

The calculations of the radiative decay rate (or the oscillator strength) have been carried out in several works, see, e.g., Refs.
~\cite{AndreaniSSC1991,CitrinPRB1993,Andreani}.
The robust theoretical calculations have been done by Iotti and Andreani~\cite{Iotti} and  D'Andrea \textit{et~al.}~\cite{Andrea},
where the general minimum of the oscillator strength for the GaAs/AlGaAs and InGaAs/GaAs QWs has been found to be at QW width $L\sim 2.5 a_{B}$.
This minimum defines the transition from the so-called weak confinement ($L \gg a_{B}$) to the strong confinement ($L < a_{B}$), when the electron and the hole are separately localized.
The experimental determination of the radiative characteristics is widespread, see, e.g., Refs.~\cite{Deveaud,Zhang,Prineas,Deveaud2005,Trifonov2015,Belykh} and references therein.
Recent measurements of the radiative decay rate have been carried out by Poltavtsev~{\em et al.}, see Refs.~\cite{Poltavtsev,Poltavtsev2014}.
In these papers, the general theoretical behavior has been experimentally confirmed, but the spread of the experimental measurements and the shortage of the available theoretical calculations motivated us to fulfill this deficiency.

In the present paper, we provide the results of numerical solution of the SE for excitons in QWs with degenerated valence band.
Partial separation (over two variables) of the center-of-mass motion and cylindrical symmetry of the problem allowed us to reduce the initial SE to the three-dimensional one.
We numerically solve the three-dimensional SE for the heavy-hole exciton in QWs of various widths and calculate the exciton binding energy and the radiative decay rate.

The numerical solution of the problem has been done for the GaAs/AlGaAs and InGaAs/GaAs QWs, which are widely experimentally and theoretically studied now as the model heterostructures~\cite{Loginov2014,Trifonov2015}.
The obtained solutions are compared with results of the variational approach with the trial function proposed in Ref.~\cite{bib5}.
The results of the numerical experiments for different widths of the QWs are discussed in detail.

We also experimentally measured the reflectance spectra for several high-quality heterostructures with InGaAs/GaAs and GaAs/AlGaAs QWs grown by molecular beam epitaxy. An analysis of exciton resonances in the spectra using the theory described in Ref.~\cite{Ivchenko} allowed us to obtain the radiative decay rates for excitons in these structures and to compare them with the numerically obtained results.

The paper is organized as follows. 
In Section 2, we derive the three-dimensional SE from the complete exciton Hamiltonian.
The direct and variational methods of numerical solution are described in Section 3.
The next section presents results on the exciton binding energy obtained by these two methods.
Section 5 contains the numerically obtained results for the radiative decay rate, their comparison with experimental data, as well as an analysis of the decay rate for the narrow and wide QWs in the framework of simple models.
The comparison of radiative decay rates and corresponding wave functions obtained by two numerical methods are discussed in Section 6.
Section 7 summarizes main results of the paper.
Additionally, some details of the numerical methods and their accuracies are given in the Appendix.

\section{Microscopic model}
The exciton in a QW is described by the SE with Hamiltonian
\begin{equation}
\label{eq1}
H = T_{e}+T_{h} -\frac{e^{2}}{\epsilon |\mathbf{r_{e}}-\mathbf{r_{h}}|}+V_{e}(\mathbf{r_{e}})+V_{h}(\mathbf{r_{h}}).
\end{equation}
Here, indices $e$ and $h$ denote the electron and the hole, respectively.
We introduce the kinetic operators for the electron, $T_{e}$, and for the hole, $T_{h}$,
the relative electron-hole distance, $|\mathbf{r_{e}}-\mathbf{r_{h}}|$, 
the electron charge, $e$, and the dielectric constant, $\epsilon$.
The square QW potential is given by
\begin{equation}
\label{eq2}
V_{j}(\mathbf{r_{j}}) = V_{j} \; \Theta \left(z^{2}_{j}-\frac{L^{2}}{4} \right),
\end{equation}
where $j=e,h$. Here $\Theta(z)$ is the Heaviside step function, $L$ is the QW width, and $V_{j}$ are the conduction- and valence-band offsets.

In the effective-mass approximation~\cite{bib1}, the kinetic operator of the electron in the conduction band is explicitly given as
\begin{equation}
\label{eq3}
T_{e} = E_{g} + \frac{\mathbf{k}^{2}_{e}}{2m_{e}},
\end{equation}
where $E_{g}$ is the energy band gap, $\mathbf{k_{e}}=-i\hbar\nabla_{e}$ is the electron momentum operator, and $m_{e}$ is the electron effective mass.
The kinetic term of the hole in the valence band is given by the Kohn-Luttinger Hamiltonian~\cite{bibLutt}, which can be written as
\begin{eqnarray}
\label{eq4}
\nonumber T_{h} &=& \frac{1}{m_{0}}
\Bigg[ \left(\gamma_{1}+\frac{5}{2} \gamma_{2} \right)\frac{\mathbf{k}_{h}^{2}}{2}
-\gamma_{2}(k^{2}_{hx}J^{2}_{x}+k^{2}_{hy}J^{2}_{y}+k^{2}_{hz}J^{2}_{z}) \\
&&-2\gamma_{3} \left( \sum_{m\ne n} \{k_{hm},k_{hn}\} \{J_{m},J_{n}\} \right) \Bigg],
\end{eqnarray}
where $\gamma_{l}$ with $l=1,2,3$ are the Luttinger parameters,
$k_{hm}$, $m=x,y,z$, are the components of the hole momentum operator,
$J_{m}$ are the $4\times 4$ angular momentum matrices~\cite{bibAlt},
$\{J_{m},J_{n}\} = (J_{m}J_{n}+J_{n}J_{m})/2$ denotes the anticommutator,
$m_{0}$ is the mass of the free electron.

We consider only the diagonal part of the expression~(\ref{eq4}), 
assuming that the nondiagonal terms contribute little to the energy states~\cite{bib1}.
Therefore, we do not consider several effects extensively discussed in literature, see, e.g., Refs.~\cite{Zimmermann,Triques}.
In particular, we ignore the heavy-hole -- light-hole coupling. On the one hand, these effects give rise to a relatively small change of the carrier and exciton energies and weak modification of the exciton-light coupling not exceeding experimental errors of the data obtained in our work. On the other hand, these simplifications allow us to get a precise numerical solution of the problem and to compare it to that obtained by the standard variational approach. 
%
%
So, up to the constant energy gap $E_{g}$, the Hamiltonian~(\ref{eq1}) has the form
\begin{eqnarray}
\label{eq5}
\nonumber H_{\text{diag}} = \frac{\mathbf{k}^{2}_{e}}{2m_{e}} +
(\gamma_{1} \pm \gamma_{2}) \frac{\left( {k}^{2}_{hx}+{k}^{2}_{hy} \right) }{2m_{0}} \\
+ (\gamma_{1} \mp 2\gamma_{2}) \frac{{k}^{2}_{hz}}{2m_{0}}
-\frac{e^{2}}{\epsilon |\mathbf{r_{e}}-\mathbf{r_{h}}|}
+V_{e}(\mathbf{r_{e}})+V_{h}(\mathbf{r_{h}}).
\end{eqnarray}
Introducing the effective masses
\begin{equation}
\label{eq51}
\begin{split}
m_{hxy} &= \frac{m_{0}}{\gamma_{1} \pm \gamma_{2}}, \\
m_{hz} &= \frac{m_{0}}{\gamma_{1} \mp 2\gamma_{2}},
\end{split}
\end{equation}
of the heavy hole (upper sign) and light hole (lower sign), 
respectively, we come to the Hamiltonian under consideration:
\begin{eqnarray}
\label{eq52}
\nonumber H_{\text{diag}} = \frac{\mathbf{k}^{2}_{e}}{2m_{e}} +
\frac{\left( {k}^{2}_{hx}+{k}^{2}_{hy} \right)}{2m_{hxy}} \\
+\frac{{k}^{2}_{hz}}{2m_{hz}}
-\frac{e^{2}}{\epsilon |\mathbf{r_{e}}-\mathbf{r_{h}}|}
+V_{e}(\mathbf{r_{e}})+V_{h}(\mathbf{r_{h}}).
\end{eqnarray}
In our study, we pay attention only to the heavy-hole excitons. The theoretical analysis for the light-hole excitons is similar and differs only in Eq.~(\ref{eq51}).

With the Hamiltonian~(\ref{eq52}), we come to the standard six-dimensional SE, $H_{\text{diag}}\Psi=E\Psi$, for the electron and the hole coupled by the Coulomb interaction~\cite{Landau}.
The translational symmetry along the QW layer allows us to reduce this equation only to the four-dimensional one by separation of the center-of-mass motion in the $(\text{x}, \text{y})$-plane.
This motion is described by an analytical part of the complete wave function, $\Psi$.
The relative motion of the electron and the hole in the exciton is described by the part $\psi(x,y,z_{e},z_{h})$ of the complete wavefunction, where $x=x_{h}-x_{e}$, $y=y_{h}-y_{e}$.
One more dimension is eliminated by taking advantage of the cylindrical symmetry of the problem and
introducing the polar coordinates $(\rho,\phi)$ for description of the relative motion, since the Coulomb potential does not depend on $\phi$.
Representing the wave function in the form
\begin{equation}
\label{eq53}
\psi(x,y,z_{e},z_{h}) = \frac{\psi(z_{e},z_{h},\rho)}{\rho} \exp{(ik_{\phi} \phi)},
\end{equation}
where $k_{\phi}=0,1,2, \ldots$, we proceed to the three-dimensional SE,
which is numerically studied in the present paper.
In Eq.~(\ref{eq53}), we introduce the factor $1/\rho$ in order to fulfill the cusp condition~\cite{Johnson} at $\rho=0$.

Since the light interacts mainly with the ground $1s$ state of the exciton, we study the case when $k_{\phi}=0$.
In this case, the SE under consideration is written as~\cite{bib4}
\begin{eqnarray}
\label{eq6}
\nonumber \Bigg( & K &
-\frac{e^{2}}{\epsilon \sqrt{\rho^{2}+(z_{e}-z_{h})^{2}}}
+V_{e}(z_{e})+V_{h}(z_{h})
\Bigg) \psi(z_{e},z_{h},\rho) \\
& = & E_{x} \psi(z_{e},z_{h},\rho)
\end{eqnarray}
where the kinetic term reads
\begin{equation}
\label{eq7}
K=-\frac{\hbar^{2}}{2m_{e}} \frac{\partial^{2}}{\partial z_{e}^{2}}
-\frac{\hbar^{2}}{2m_{hz}} \frac{\partial^{2}}{\partial z_{h}^{2}}
-\frac{\hbar^{2}}{2\mu}
\left(\frac{\partial^{2}}{\partial \rho^{2}} - \frac{1}{\rho} \frac{\partial}{\partial \rho} +\frac{1}{\rho^{2}} \right),
\end{equation}
and $\mu=m_{e} m_{hxy}/\left( m_{e} + m_{hxy} \right)$ is the reduced mass in the $(\text{x}, \text{y})$-plane.
The obtained three-dimensional SE~(\ref{eq6}) cannot be solved analytically
because the potential terms do not allow further separation of the variables.
The approximate solution is also difficult for the finite QW potentials $V_{e,h}$ due to penetration of the wave function under the barriers.

In our study, SE~(\ref{eq6}) is solved numerically
and the energy and corresponding wave function are obtained for QWs of various widths and compositions of the QW layer and barriers.
Other parts of the complete wave function are analytically known, though they are not of the practical importance.

\begin{table*}[htbp!]
\caption{
Material parameters used for solving the eigenvalue problem~(\ref{eq6}). Energy gap mismatch denoted as $\Delta E_g$ is calculated for AlGaAs heterostructure based on the data from Ref.~\cite{bib5}, for InGaAs, it is based on Ref.~\cite{Andrea}. Right side of the table contains the effective masses for pure materials. Masses for the ternary alloys are obtained using the linear interpolation on $x$. Material parameters for AlGaAs and InGaAs heterostructures are listed here according to Ref.~\cite{bib5} and Ref.~\cite{bib9}, respectively.}
\label{t1}
\begin{ruledtabular}
\begin{tabular}{cccccccccc}
 Heterostructure & $x$ & $\epsilon$ & $V_{e}/V_{h}$ & $\Delta E_{g}$~(meV) & & $m_{e}/m_0$ & $m_{hz}/m_0$ & $m_{hxy}/m_0$ & $\mu/m_0$   \\
\hline
\multirow{2}{*}{GaAs/Al$_x$Ga$_{1-x}$As} & \multirow{2}{*}{0.3} & \multirow{2}{*}{$12.53$} & \multirow{2}{*}{$0.65/0.35$} & \multirow{2}{*}{$365.5$} & GaAs & $0.067$ & $0.377$ & $0.111$ & $0.042$ \\
&  &  &  &  & AlAs & $0.150$ & $0.478$ & $0.242$ & $0.093$\\
\hline
\multirow{2}{*}{In$_x$Ga$_{1-x}$As/GaAs} & 0.02 &\multirow{2}{*}{$12.53$} & $0.65/0.35$ & $30$ & GaAs & $0.067$ & $0.350$ & $0.111$ & $0.042$\\
 & 0.09 & & $0.55/0.45$ & $139$ & InAs & $0.026$ & $0.333$ & $0.035$ & $0.015$\\
\end{tabular}
\end{ruledtabular}
\end{table*}

\section{Numerical methods}
%
We performed the direct numerical solution of Eq.~(\ref{eq6})
for precise calculations of the exciton ground state energy, $E_{x}$.
The exponential decrease of the wave function at large values of variables allows us to impose the zero boundary conditions for the wave function at the boundary of some rectangular domain.
The size of this domain varies from dozens QW widths (for small widths) down to several QW widths (for large widths).
Therefore, the studied boundary value problem (BVP) is formed by Eq.~(\ref{eq6}) and the zero boundary conditions at $\rho=0$, some large value of $\rho$ and at large positive and negative values of the variables $z_{e,h}$.
Since SE~(\ref{eq6}) is the standard three-dimensional partial differential equation of elliptic type, the direct numerical solution of the BVP is feasible using available computational facilities.
For this purpose, we employed the fourth-order finite-difference approximation of the derivatives on the equidistant grids over three variables.
The precise values of the studied quantities are obtained by the extrapolation of the results of calculations as the grid step goes to zero.
The details on the numerical scheme and theoretical uncertainties are described in the Appendix.

The nonzero solution of the homogeneous equation~(\ref{eq6}) with trivial boundary conditions can be obtained by the diagonalization of the matrix constructed from the Hamiltonian of Eq.~(\ref{eq6}).
The obtained square matrix is large, nonsymmetric and sparse. The typical size of the matrix is of the order of $10^{6}$, so
we keep in the calculations only nontrivial matrix elements of a few diagonals.
The diagonalization of such a matrix is difficult, however, a small part of the spectrum can be easily obtained.
Using the Arnoldi algorithm~\cite{bib6}, we have calculated the lowest eigenvalue of the matrix and the corresponding eigenfunction.
As a result, the ground state energy, $E_{x}$, and the corresponding wave function
have been obtained for various widths of QW.

One of the alternative numerical methods is the approximate solution of the three-dimensional SE~(\ref{eq6}) by the variational approach. 
This technique has been applied by many authors, see Refs.~\cite{bib4,Greene,bib5,Ivchenko}.
In the framework of the variational approach, the ground state energy of the system with the Hamiltonian $H_{x}$ is determined by the minimization of the functional
\begin{equation}
\label{v1}
F = \frac{\left \langle \psi | H_{x} | \psi  \right \rangle}{\left \langle \psi |  \psi  \right \rangle},
\end{equation}
with respect to some free parameters of the trial wave function $\psi$.
The Hamiltonian $H_{x}$ corresponds to the left-hand side part of Eq.~(\ref{eq6}).
For numerical calculation of the integrals in Eq.~(\ref{v1}) one has to define the trial function.
In Ref.~\cite{bib5}, the trial wave function having the form
\begin{equation}
\label{v2}
\psi(z_{e},z_{h},\rho) = \psi_{e}(z_{e}) \psi_{h}(z_{h}) \exp \left(-\frac{\alpha}{a_{B}}\sqrt{\rho^{2}+\lambda(z_{e}-z_{h})^{2}} \right)
\end{equation}
is applied.
Here, the functions $\psi_{e}(z_{e})$ and $\psi_{h}(z_{h})$ are the ground state wave functions of the free electron and free hole, respectively,
and $\alpha$, $\lambda$ are the varying parameters.

The shortcoming of the described variational approach is that the trial wave function has the prescribed form.
Moreover, it assumes the partial separation of the variables, whereas the Coulomb potential in Eq.~(\ref{eq6}) does not allow that separation.
Of course, one can define even more complicated trial functions~\cite{Greene,Schiumarini},
but asymptotically (for small and large widths of QW) they have to be reduced to the function~(\ref{v2}).
This fact as well as the numerical simplicity of this anzatz provoked many authors to use it for calculation of the exciton binding energy, $R_{x}$.
However, the accuracy of the results has not been studied in detail so far.
The obtained wave function has also not been applied to calculate the exciton radiative characteristics.

We apply the described numerical algorithms for solving the SE with parameters for the GaAs/Al$_{x}$Ga$_{1-x}$As and In$_{x}$Ga$_{1-x}$As/GaAs QWs,
which are widespread in the contemporary experimental studies.
The parameters are general for such types of the QWs and, in particular, simulate the typical ratio of the band offsets at the GaAs/Al$_{0.3}$Ga$_{0.7}$As interface: $V_{e}/V_{h}\approx2$.
In the calculation, we have used the heavy-hole Al$_{x}$Ga$_{1-x}$As parameters reported in Ref.~\cite{bib5} as well as the In$_{x}$Ga$_{1-x}$As ones in Ref.~\cite{bib9}.
They are presented in Table~\ref{t1}.
Parameter
$x$ denotes the concentration of Al (In) in the Al(In)GaAs solid solutions, $\Delta E_{g}(x)$ is the difference of the energy gap $E_{g}(x)$ for $x=0$ and $x > 0$.
For InGaAs, ratios of potential barriers were taken $V_{e}/V_{h}=65/35$ for concentration $x=0.02$ and $V_{e}/V_{h}=55/45$ for $x=0.09$.
For simplicity, we neglect the difference of the electron and hole masses in the QW and in the barrier layers.
We ignore the small discontinuity of the dielectric constant at the QW interfaces as well~\cite{Kumagai}.
%

\section{Exciton binding energy}
%
The exciton binding energy, $R_{x}$, is defined by the exciton ground state energy, $E_{x}$,
with respect to the quantum confinement energy of the electron, $E_{e}$, and the hole, $E_{h}$,
in QW. This relation is given by the formula
$$
R_{x}=E_{e}+E_{h}-E_{x},
$$
which characterizes the energy of the Coulomb interaction of the electron and the hole trapped in the QW potentials $V_{e}$ and $V_{h}$, respectively.
Energies $E_{e}$ and $E_{h}$ are obtained from solution of the corresponding one-dimensional SEs for the electron and the hole in QW~\cite{bib7}.

\begin{figure}[htbp!]%
\includegraphics*[width=\linewidth]{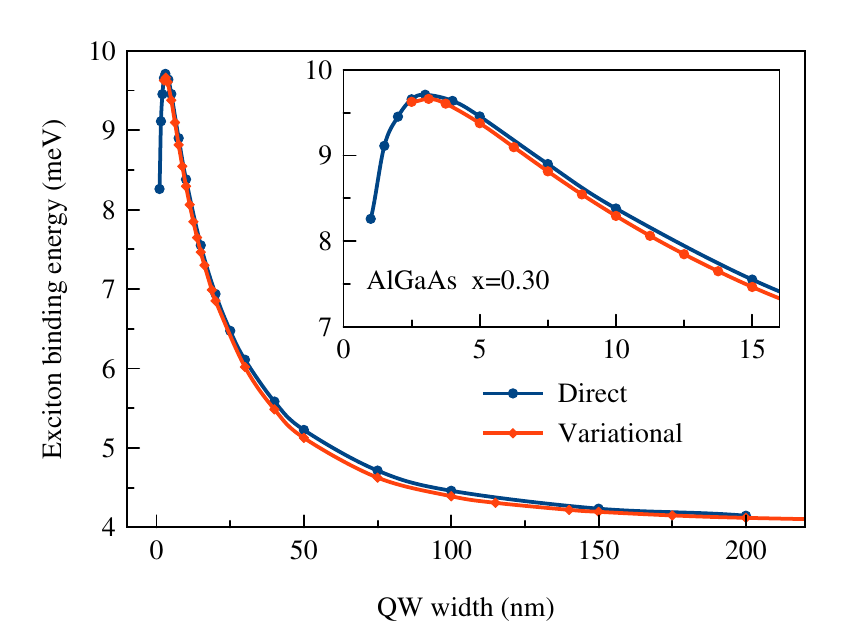}
\caption{%
Exciton binding energy, $R_{x}$, obtained by the direct numerical solution of Eq.~(\ref{eq6}) (blue curve) and by the variational approach (red curve) for GaAs/Al$_{0.3}$Ga$_{0.7}$As QWs as a function of the QW width.
}
\label{fig2}
\end{figure}

The direct microscopic calculations of the exciton binding energy have been carried out for various widths of the GaAs/Al$_{0.3}$Ga$_{0.7}$As QW.
The concentration of aluminum in GaAs has been chosen to be $x=0.3$ in accordance to the concentration of the grown samples.
Fig.~\ref{fig2} shows the exciton binding energy, $R_{x}$, obtained from the microscopic calculations and the variational approach as a function of the QW thickness.
Surprisingly, the results of two methods coincide with the precision of $0.1$~meV for QW widths $L \ge 2.5$~nm.
This result means that the trial function~(\ref{v2}) is a good approximation of the exciton wave function for the lowest state.
The variational approach gives a bit smaller binding energy, which is the expectable result because the minimum of functional~(\ref{v1}) is achieved for exact exciton wave function rather than for the trial function.

The uncertainties of the microscopic calculations of $R_{x}$ come from the calculated values of $E_{x}$ because the quantum confinement energies, $E_{e}$ and $E_{h}$, can be calculated with an arbitrary precision.
We obtained the relative uncertainty of $R_{x}$ to be smaller than $1\%$ for $1.0 \le L < 2.5$~nm and decreasing down to values smaller than $0.1\%$ for wider QWs.
Although the uncertainty is much smaller than the difference of the results of about $0.1$~meV obtained by two numerical methods, this discrepancy seems to be negligible for comparison with contemporary experimental data.
Therefore, both methods can be successfully used for determination of the exciton binding energy.

Interestingly enough, the maximum binding energy of $9.71$~meV is achieved by the direct microscopic solution at the QW width of about $3$~nm.
The variational approach gives for this QW $E_{x}=9.66$~meV at the optimal parameters of the trial function~(\ref{v2}) are $\alpha=1.438$, $\lambda=0.3125$.
In contrast to the case of QWs with infinitely high barriers, for this case, the binding energy decreases for thinner QWs due to the penetration of the carriers into the barriers.
In the other limit of very thick QWs, we obtained the values approaching the free exciton binding energy for the bulk crystal.
From Fig.~\ref{fig2}, one can see that, for our case, this energy is $R_{x}=4.10\pm0.03$~meV that is slightly less than reported in Ref.~\cite{bib1}. 
For the variational approach, the parameters of the trial function~(\ref{v2}) for the QW of width $L=200$~nm, where the exciton can be treated as the free one, are $\alpha=1.0604$, $\lambda=1.1858$. These values are close to the expectable ones for the bulk crystal: $\alpha_{\infty}=1$, $\lambda_{\infty}=1$~\cite{Ivchenko}.
The overall behavior of the binding energy for QW widths $1 \le L \le 150$~nm can be approximated with the accuracy of $0.07$~meV by simple formula $R_{x}(L)=(3.5431L^{2}+146.694L)/(L^{2}+11.1758L+6.0473)$.

\begin{figure}[htbp!]%
\begin{tabular}{c}
\begin{minipage}{1.0\linewidth}
\includegraphics[width=\linewidth]{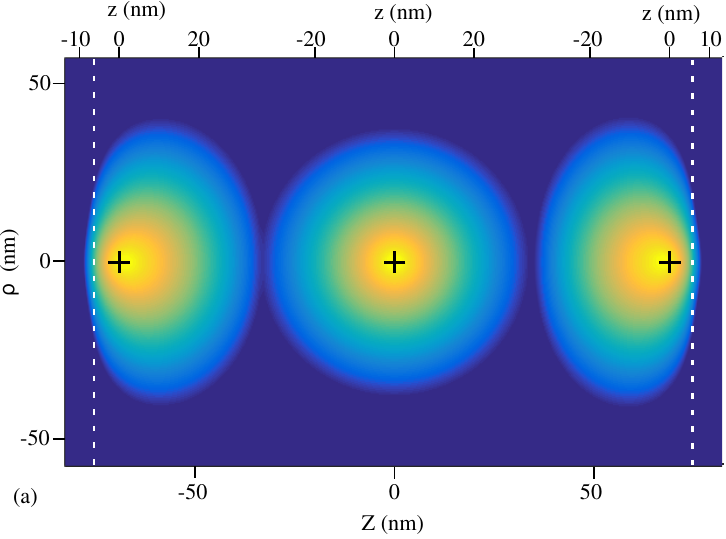}
\end{minipage}
\\
\begin{minipage}{1.0\linewidth}
\includegraphics[width=\linewidth]{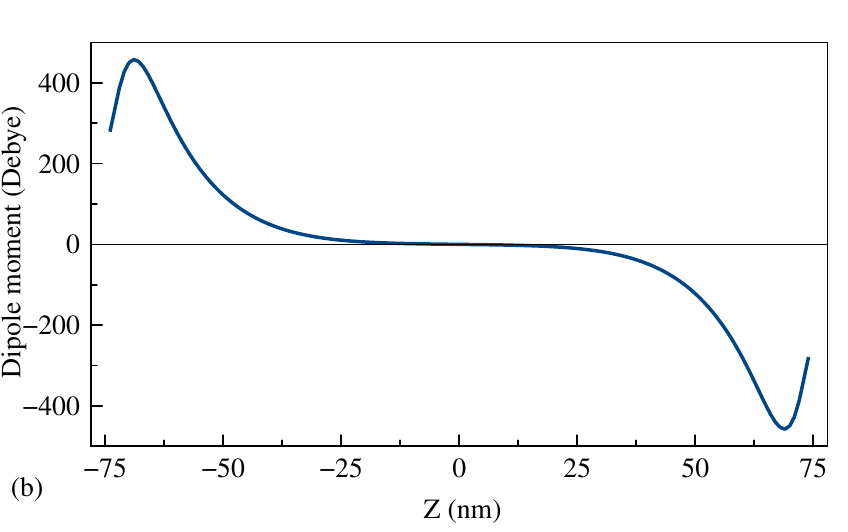}
\end{minipage}
\end{tabular}
\caption{%
\label{wf1}
(a) The slices of the exciton wave function squared, $|\psi(Z,z,\rho)|^{2}$, as functions of relative coordinates $(z,\rho)$ for three different center-of-mass coordinates $Z=-69,0,69$~nm of the exciton.
The QW width, $L$, equals to 150~nm.
The coordinates of the center-of-masses of the excitons are depicted by small crosses.
The QW interfaces are drawn by the dashed vertical lines.
The probability for the electron and the hole to be at distance $r = \sqrt{z^2 +\rho^2}$ is shown by the false colors in the logarithmic scale.
(b) The exciton dipole moment as a function of the center-of-mass coordinate.
}
\end{figure}

Together with the exciton ground state energy, in the direct microscopic calculation we have obtained the corresponding wave function.
In the wide QWs, the wave function $\psi(z_{e},z_{h},\rho)$ can be presented as a function of relative coordinate, $z=z_{e}-z_{h}$, and center-of-mass coordinate, $Z=(m_{e}z_{e}+m_{h}z_{h})/(m_{e}+m_{h})$.
The slices of $|\psi(Z,z,\rho)|^{2}$ as functions of $z$ and $\rho$ for three different center-of-mass coordinates $Z$ are shown in Fig.~\ref{wf1}(a) for the QW width $ L = 150$~nm.
These slices show the probability distribution for relative distance between the electron and the hole in the exciton.
The center plot shows the exciton localized in the center of the QW while the side plots present the exciton near the QW interfaces.
The magnitude of $|\psi(Z,z,\rho)|^{2}$ for the side plots is in several orders smaller than for the center plot.

As seen from Fig.~\ref{wf1}(a), the probability distribution at $Z=0$~nm is spherically symmetric and reveals some distortion near the QW interfaces. 
The exciton in a wide QW can be considered as slowly propagating across the QW that is its kinetic energy is small compared to the Coulomb energy. Therefore an adiabatic approximation can be used, when the electron-hole relative motion in the exciton is considered to be much faster than the center-of-mass motion. In the framework of this approximation, the obtained distortion can be treated as an appearance of a static dipole moment of the exciton when it encounters the barrier.

The exciton static dipole moment has been extensively studied for indirect excitons in coupled QWs, see, e.g., Refs.~\cite{Butov-JPhys2004,Laikhtman-PRB2009,Stern-Science2014} and references therein.
In our case of the direct exciton, it has been calculated using the standard definition: $D_{x} = e\left \langle \psi|z|\psi \right \rangle$.
The dependence of the dipole moment on the exciton position in the QW is presented in Fig.~\ref{wf1} (b).
It is clearly seen that the maximum dipole moment is achieved near the barrier of the QW,
where the wave function undergoes the maximum distortion.
When the wave function is spherically symmetrical ($Z=0$), the dipole moment is zero and increases in absolute value as the exciton approaches the barriers.
If the exciton is very close to the barrier, the wave function penetrates under the barrier, distortion diminishes and the dipole moment decreases.

\section{Exciton-light coupling}

\subsection{Radiative decay rate}
\label{radiative_rate}

Exciton-light coupling is usually characterized by either the radiative decay rate or the oscillator strength~\cite{Knox,AndreaniSSC1991,CitrinPRB1993,Ivchenko}. 
The radiative decay rate, $\Gamma_0$, characterizes the decay of electromagnetic field emitted by the exciton after the pulsed excitation: $E(t) = E(0)\exp(-\Gamma_0 t)$. 
A consistent exciton-light coupling theory is presented, e.g., in the monograph of Ivchenko~\cite{Ivchenko}. It provides the folowing expression for $\Gamma_0$:
\begin{equation}
\label{Gamma0}
\Gamma_{0} =\frac{2 \pi q}{ \hbar \epsilon} \left(\frac{e|p_{cv}|}{m_{0} \omega_{0}} \right)^{2}
\left | \int\limits_{-\infty}^{\infty} \Phi(z) \exp(iqz) dz  \right |^{2},
\end{equation}
where $q=\sqrt{\epsilon}\omega/c$ is the light wave vector, $\omega_0$ is the exciton frequency, $|p_{cv}|=\left \langle u_{v} | \epsilon \cdot \mathbf{p} | u_{c} \right \rangle$ is the matrix element of the momentum operator between the single-electron conduction- and valence-band states, and $\Phi(z)\equiv\psi(z=z_{e}=z_{h},\rho=0)$.

The simplification of Eq.~(\ref{Gamma0}) used in Refs.~\cite{Iotti,CitrinPRB1993} takes into account that the QW width is much smaller than the light-wave length $2\pi/q$. It allows one to replace $\exp(iqz)$ in the integral by unity.
In this case, the radiative decay rate $\Gamma_{0}$ is closely related to the oscillator strength per unit area~\cite{CitrinPRB1993}, $f/S$, by the formula
\begin{equation}
\label{eq9}
\Gamma_{0} = \frac{\pi e^{2}}{n m_{0}c} \left( \frac{f}{S} \right),
\end{equation}
where $n$ is the refraction coefficient ($\sqrt{\epsilon}=n+ik$).

The wave function obtained from the microscopic calculation allowed us to calculate the radiative decay rate $\Gamma_{0}$ of the exciton ground state according to Eq.~(\ref{Gamma0}).
We calculated $\Gamma_{0}$ for GaAs/Al$_{0.3}$Ga$_{0.7}$As QWs of various widths from 1~nm to 300~nm.
In the calculations, $|p_{cv}|^{2}=m_{0}E_{p}/2$, where $E_{p}=28.8$~eV for GaAs and $E_{p}=21.5$~eV for InAs are taken from Ref.~\cite{bib9}. The exciton frequency $\omega_0$ is calculated using bandgap $E_g = 1.520$~eV for GaAs~\cite{Ivchenko} and parameters listed in Tab.~\ref{t1}.

\begin{figure}[htbp!]%
\includegraphics*[width=\linewidth]{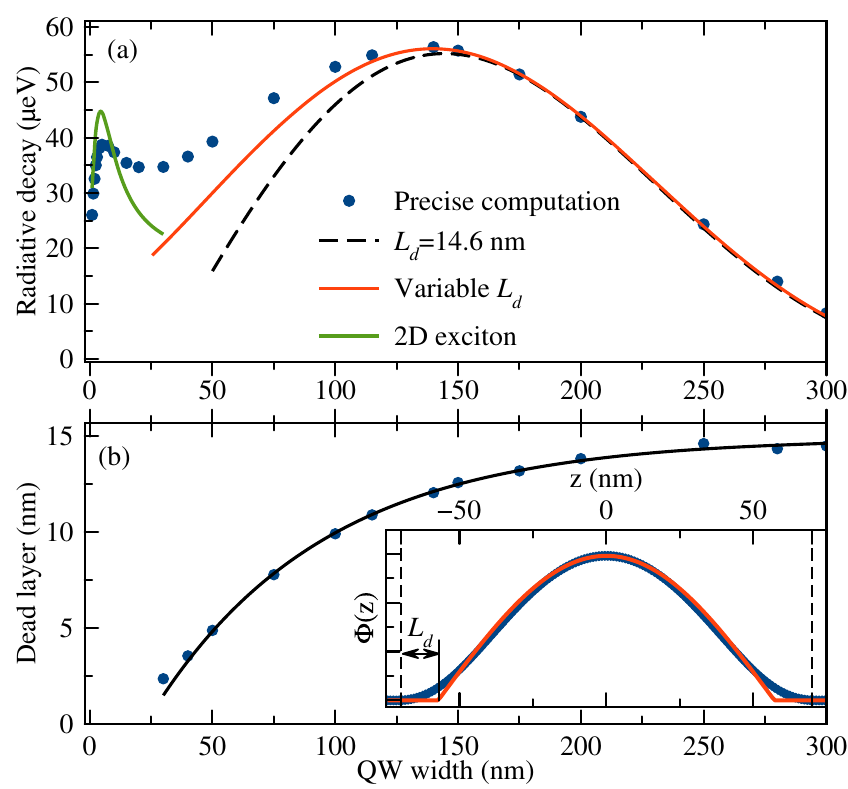}
\includegraphics*[width=\linewidth]{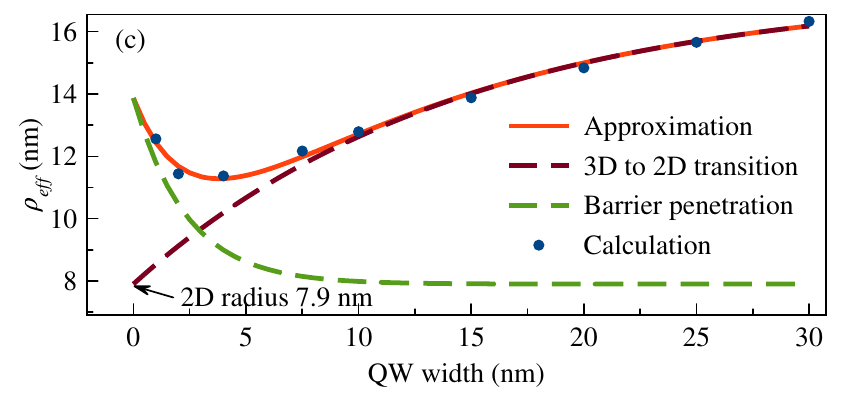}
\caption{%
(a) Radiative decay rate in energy units, $\hbar\Gamma_{0}$, versus QW width, $L$. Blue dots are obtained in
the precise microscopic calculation. Dashed line corresponds to $\Gamma_0$ for the bulk exciton in a QW with infinite barriers and constant dead layer $L_{d} = 14.6$~nm. Solid red line is the calculation with the variable dead layer shown in (b).
The green solid line corresponds to $\Gamma_0$ in the 2D exciton model with the varying effective radius $\rho_{\text{eff}}$ shown in (c).
(b) The variable dead layer as a function of QW width.
The inset: function $\Phi(z)$ and its approximation by $\cos(\pi Z/L^*)$ for QW with $L = 150$~nm. The two-side arrow marked ``$L_d$'' illustrates definition of the dead layer.
(c) Effective 2D exciton radius, $\rho_{\text{eff}}$, extracted form the calculations (solid points) for the QW widths 0 -- 30~nm and its approximation by function~(\ref{2Drhofit}) (solid curve) with parameters:  $C_1=9.5\pm0.4$~nm, $L_1=14\pm1$~nm, $C_2=6.0\pm0.5$~nm, $L_2=2.4\pm0.3$~nm. Dashed curves show contributions $f_{\text{squ}}(L)$ and $f_{\text{pen}}(L)$ in Eq.~(\ref{2Drhofit}).
}
\label{OscillatorStrength}
\end{figure}

Fig.~\ref{OscillatorStrength} shows the radiative decay rate in energy units, $\hbar\Gamma_{0}$, as a function of the QW width.
The radiative decay rate reaches its maximum at the QW width of about $130$~nm, that approximately corresponds to the half of the light wavelength in the QW material, $\lambda (\text{GaAs}) = \lambda (vac)/n(\text{GaAs}) = 230$~nm, where $n(\text{GaAs}) = 3.6$ is the refractive coefficient of GaAs at the photon energy $\hbar\omega = E_g(\text{GaAs})$. So, this maximum of $\Gamma_0$ corresponds to the maximal overlap of the exciton wave function $\Phi(z)$ and the light wave, see Eq.~(\ref{Gamma0}).

As the QW width decreases, $\Gamma_{0}$ also decreases due to the diminish of the overlap integral in Eq.~(\ref{Gamma0}). For small QW widths, however, $\Gamma_0$ grows that correlates with the increase of the exciton binding energy (compare with Fig.~\ref{fig2}).
Therefore, we may suppose that this maximum of $\Gamma_0$ is caused by squeezing of the exciton in the narrow QWs by the QW potential.
The squeezing gives rise to increase of the probability to find electron and the hole in the same position ($z_e = z_h$ and $\rho = 0$). Between these maxima of $\Gamma_0$ there is a minimum for the QW width of about 30~nm, which corresponds to the exciton Bohr diameter. The presence of such minimum was earlier pointed out by Iotti and Andreani~\cite{Iotti}.

We compared our results with two simple approximations: the exciton in the bulk semiconductor for wide QWs and two-dimensional approximation for narrow QWs.
Both the approximations are shown in Fig.~\ref{OscillatorStrength} by various curves.
The wave function of the exciton in the bulk semiconductor is given as~\cite{Ivchenko}
\begin{equation}
\psi(Z,r)=\frac{1}{\sqrt{\pi a_{B}^{3} L^{*}}}\cos\left(\frac{\pi}{L^{*}}Z\right) \exp {\left({-\frac{r}{a_{B}}}\right)},
\label{psiWide}
\end{equation}
where $r$ is the electron-hole distance, $L^{*}=L-2 L_{d}$ is the effective QW width obtained from the QW width $L$ and the dead layer $L_{d}$~\cite{DAndrea1982,exp2,Ubyivovk}.
The part of the wave function~(\ref{psiWide}) depending on $Z$ is represented only by function $\cos{(\pi Z /L^{*})}$.
It leads us to the conventional definition of the dead layer, which is the distance from the QW interface to the point where the cosine approximation of the exciton wave function becomes zero.
This definition of the dead layer is illustrated by the inset of Fig.~\ref{OscillatorStrength} (b).
Substituting the wave function~(\ref{psiWide}) in Eq.~(\ref{Gamma0}), one can calculate the radiative decay rate in a QW with infinite barriers and constant dead layer $L_{d} = 14.6$~nm.
It is shown by dashed curve in Fig.~\ref{OscillatorStrength} (a).
One can see that this approximation is acceptable for the QW widths $L \ge 140$~nm.

A better approximation of the dependence $\Gamma_0(L)$ can be achieved using the variable dead layer~\cite{Schiumarini}.
We extracted the variable dead layer by fitting the function~(\ref{psiWide}) at $r=0$ to the numerically obtained $\Phi(z)$ [see inset in Fig.~\ref{OscillatorStrength}~(b)].
Extracted values of $L_d$ are shown in Fig.~\ref{OscillatorStrength} (b) by solid points.
Dependence of $L_d$ on the QW width $L$ can be well approximated by the phenomenological formula: $L_{d}=a(1-\exp{(-L/L_{0})})+b$, where $a=20.6\pm0.5$~nm, $L_{0}=70\pm2$~nm, and $b=-5.7\pm0.7$~nm.
Using this approach and $a_{B}=14.55$~nm, we obtained more accurate approximation of $\hbar\Gamma_{0}(L)$ which is shown in Fig.~\ref{OscillatorStrength}~(a) by red solid curve.
This approximation is appropriate for the QW widths down to $100$~nm.

For narrower QWs, both the bulk exciton wave function and the idea of variable dead layer are no longer applicable.
Instead, for thin QWs, we implemented the two-dimensional (2D) exciton approximation.
The wave function of the 2D exciton has the form~\cite{Ivchenko}
\begin{equation}
	\psi(z_e,z_h,\rho)=\psi_{e}(z_e)\psi_{z}(z_h)\sqrt{\frac{2}{\pi \rho^{2}_{\text{eff}}}}\exp{\left(-\frac{\rho}{\rho_{\text{eff}}}\right)},
\label{psiThin}
\end{equation}
where $\rho_{\text{eff}}$ is the effective 2D exciton radius, $\psi_{e}(z_{e})$ and $\psi_{h}(z_{h})$ are wave functions of the free electron and free hole in a QW with finite barriers. Function~(\ref{psiThin}) is the solution of eigenvalue problem~(\ref{eq6}) with 2D Coulomb potential, $-e^2/(\epsilon \rho)$.
In the true 2D exciton problem, $\rho_{\text{eff}}$ is the 2D Bohr radius
\begin{equation}
	\rho_{2\text{D}}=\frac{\hbar^2 \epsilon}{2 \mu e^2},
\label{2Drhoideal}
\end{equation}
which is twice smaller than $a_{B}$.

The exciton in the QWs with finite heights of the barriers does not reach two-dimensional limit due to penetration of the exciton wave function into the barriers.
Therefore, we consider $\rho_{\text{eff}}$ as a characteristic parameter. 
We fitted the wave function~(\ref{psiThin}) to the numerically obtained wave function $\Phi(z)$ by varying $\rho_{\text{eff}}$. Extracted effective 2D exciton radius values are presented in Fig.~{\ref{OscillatorStrength}} (c).
It is seen that $\rho_{\text{eff}}$ decreases as $L \to 4$~nm and then increases again as $L \to 0$.
This behavior is well described by a phenomenological function
\begin{eqnarray}
\nonumber	f(L) &=& f_{\text{squ}}(L) + f_{\text{pen}}(L) \\ 
                     &=& \left[\rho_{2\text{D}} + C_1 \left(1 - e^{-L/L_1}\right)\right] 
               	+ \left[C_2 e^{-L/L_2}\right].
	\label{2Drhofit}
\end{eqnarray}
The first part of this function, $f_{\text{squ}}(L)$, decreases with $L \rightarrow 0$ down to the 2D-limit given by Eq.~(\ref{2Drhoideal}), $\rho_{2\text{D}} = 7.9$~nm for GaAs, and reflects the squeezing of exciton in narrow QWs.
Another one, $f_{\text{pen}}(L)$, increases as $L \to 0$ and reflects the penetration of exciton into the barrier.
Using this dependency, we calculated $\hbar \Gamma_0$ for the 2D exciton model, see Fig.~\ref{OscillatorStrength} (a).
As it is seen, this model, even with varying $\rho_{\text{eff}}$, is adequate in terms of the radiative decay rate only for narrow QWs $L \le 15$~nm.

The described approximations are applicable for narrow or wide QWs separately.
The excitons in QWs of intermediate widths are not described by these models.
Only the direct microscopic calculation provides the precise values of the radiative decay rate for the wide range of QW widths.

\subsection{Experimental determination of $\Gamma_0$}

Exciton reflectance spectra can be used to obtain radiative decay rate.
According to the theory summarized in Ref.~\cite{Ivchenko}, the amplitude reflectance coefficient of the QW with exciton resonance is given as
\begin{equation}
\label{reflection}
r_{QW}=\frac{i\Gamma_{0}}{\tilde{\omega}_{0}-\omega-i(\Gamma+\Gamma_{0})},
\end{equation}
where $\tilde{\omega}_{0}$ is the renormalized exciton resonance frequency and $\omega$ is the frequency of the incident light.
The nonradiative decay rate $\Gamma$ in Eq.~(\ref{reflection}) takes into account an additional broadening of the resonances due to nonradiative processes.
Reflectance $r_{QW}$ is strictly related to the reflectance coefficient of the whole heterostructure, $R$, which, in turn, can be measured in experiment.

For the single QW heterostructure the relation between $R$ and $r_{\text{QW}}$ can be found in Ref.~\cite{Ivchenko}.
In a structure with several QWs as those used in our study, the reflectance coefficient is generalized to
\begin{equation}
	R=\left|\frac{r+\sum_j{r_{\text{QW}j}e^{i\phi_j}}}{1+r \sum_j{r_{\text{QW}j}e^{i\phi_j}}}\right|^2.
	\label{multistructrefl}
\end{equation}
Here $r$ is the Fresnel reflectance coefficient from the surface of a heterostructure and $\phi$ is the phase shift of the light wave reflected by the QW with respect to that reflected by the structure surface.
\begin{figure}[htbp!]%
\begin{tabular}{c}
\begin{minipage}{1.0\linewidth}
\includegraphics[width=\linewidth]{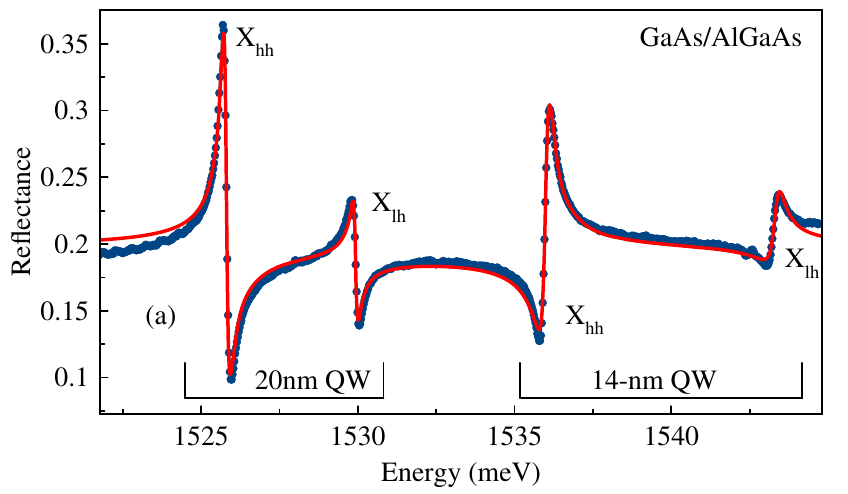}
\end{minipage}
\vspace{0.1cm}\\
\begin{minipage}{1.0\linewidth}
\includegraphics[width=\linewidth]{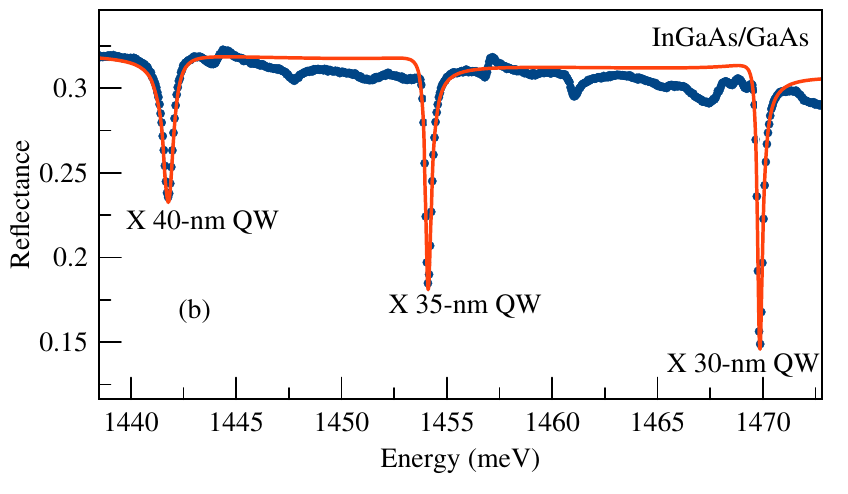}
\end{minipage}
\end{tabular}
\caption{%
\label{fig4exp}
(a) The experimental reflectance spectrum (blue curve) of the heterostructure with GaAs/AlGaAs QWs as well as the fit (red curve) of the spectrum. The ground heavy- and light-hole exciton states are denoted by $X_{hh}$ and $X_{lh}$, respectively. (b) The same is for the heterostructure with InGaAs/GaAs QWs.
}
\end{figure}
Effectively, Eq.~(\ref{multistructrefl}) is the direct relation between the reflectance $R$ and the radiative decay rate $\Gamma_0$.

For comparison of the theoretical modeling with experimental results, several high-quality heterostructures with QWs grown by molecular beam epitaxy have been selected for study of their reflectance spectra. The spectra were measured using a simple setup consisting of a white light source, a cryostat, and a spectrometer equipped with a CCD camera. Special precautions have been taken to accurately calibrate the absolute value of reflectance. For this purpose, a monochromatic light of a continuous-wave titanium-sapphire laser was used to measure the reflectance at a spectral point beyond the exciton resonances.

Fig.~\ref{fig4exp} demonstrates examples of the reflectance spectra (blue curves) for the GaAs/AlGaAs and InGaAs/GaAs heterostructures.
We used the Eq.~(\ref{multistructrefl}) to fit the spectra with $\Gamma_0$, $\Gamma$, $\omega_0$ and $\phi$ as the fitting parameters for each resonance (red curves).
The radiative decay rates in energy units, $\hbar\Gamma_0$, obtained for series of QW heterostructures are presented in Tab.~\ref{tableexp}.
We fitted only the ground heavy-hole and light-hole exciton states.
In the InGaAs spectra, we were able to reliably determine only the heavy-hole exciton resonances because the light-hole ones are significantly detached by the strain~\cite{Van_de_Walle}.
\begin{table}[htbp!]
\begin{ruledtabular}
\caption{Radiative decay rate $\hbar\Gamma_0$ in energy units ($\mu$eV) together with the fit standard deviations. The data were extracted from the measured spectra of samples with QWs of various widths.\label{tableexp}}
\begin{tabular}{ c c c c c c }
Structure & QW width (nm) & $\hbar\Gamma_0$ of $X_{hh}$ & $\hbar\Gamma_0$ of $X_{lh}$ \\
\hline
GaAs/AlGaAs & 14 & $36.6\pm0.6$ & $14.9\pm0.7$ \\
& 20 & $35.4\pm0.4$ & $10.9\pm0.4$ \\
\hline
InGaAs/GaAs & 2 & $30.0\pm0.2$ & \\
& 3 & $27.3\pm0.5$ & \\
& 3 & $25.8\pm0.6$ & \\
& 30 & $38.2\pm0.8$ & \\
& 35 & $35.1\pm0.6$ & \\
& 40 & $36.8\pm0.4$ & \\
& 95 & $58.0\pm0.5$ & \\
\end{tabular}
\end{ruledtabular}
\end{table}
The shape of exciton peculiarities is known to be defined by phase $\phi_j$ in Eq.~(\ref{multistructrefl}) determined by the QW-to-surface distance in real sample.
Relatively small peculiarities of the spectrum observed in Fig.~\ref{fig4exp}(b) are, most probably, the excited quantum confined exciton states, which are beyond the scope of the present paper.

\begin{figure}[htbp!]%
\includegraphics*[width=\linewidth]{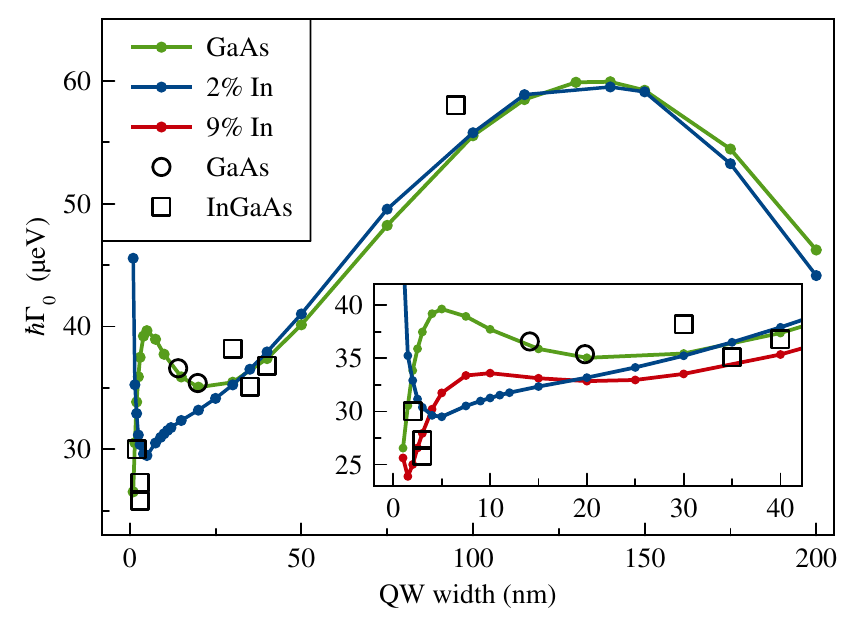}
\caption{%
\label{TheoryExp}
Radiative decay rate for In$_{x}$Ga$_{1-x}$As/GaAs QWs with 2\% and 9\% of In (blue and red curves, respectively) as well as for GaAs/AlGaAs QWs (green curve).
The empty squares and circles show the experimental data for InGaAs/GaAs and GaAs/AlGaAs QWs, respectively.}
\end{figure}

Experimentally obtained radiative decay rates for heavy-hole excitons listed in Tab.~\ref{tableexp} are compared to the calculated data in Fig.~{\ref{TheoryExp}}. 
In the calculations, we simulated GaAs/Al$_{0.3}$Ga$_{0.7}$As and In$_{0.02}$Ga$_{0.98}$As/GaAs heterostructures with 365.5~meV and 30~meV band offsets (see Tab.~\ref{t1}), respectively.
The calculations indicate that the radiative decay rate for GaAs/AlGaAs heterostructure grows as $L$ diminishes in the range $L = 5-20$~nm.
The experimentally studied GaAs/AlGaAs structures exhibit perfect agreement with the calculated results. 
In particular, they support the
tendency of increase of $\hbar \Gamma_0$ with decrease of the QW width in the range $L = 5-20$~nm.
This tendency is also confirmed by many experimental data in Ref.~\cite{Poltavtsev}.

Experimental results for InGaAs/GaAs heterostructures demonstrate a trend of $\hbar \Gamma_0$ to diminish with decrease the QW width down to $L \approx 5$~nm, which is in agreement with the calculations. 
At the same time, the experimental data are more spread around values expected from computation.
Large spread of the data is also observed in Ref.~\cite{Poltavtsev2014}. 
We explain this spread mainly by the indium concentration variation from one sample to another in the range 2 -- 5\%.
 We should note that high mobility of indium atoms during the growth process makes InGaAs heterostructures less predictable as compared to AlGaAs ones, in particular, due to the segregation effect~\cite{Van_de_Walle}.

\subsection{$\hbar \Gamma_0$ in shallow QWs}

The large difference in behavior of the radiative decay rate for GaAs/AlGaAs and InGaAs/GaAs in the range of small QW widths requires a particular analysis. 
We believe that this difference is related to different depth of the QWs.
To check this assumption, we carried out computations of $\hbar\Gamma_0$ for different concentrations $x$ in the  In$_{x}$Ga$_{1-x}$As/GaAs heterostructures that results in different heights of the QW barriers.
In particular, for $x=0.02$ and $0.09$ the barriers are $V_{e}=19.5$~meV and $V_{e}=76.5$~meV, respectively.

In the inset of Fig.~{\ref{TheoryExp}}, the radiative decay rates for In$_x$Ga$_{1-x}$As/GaAs heterostructures with $x=0.02$ and $x=0.09$ and for GaAs/Al$_{0.3}$Ga$_{0.7}$As heterostructures are shown.
It demonstrates the evolution of the radiative decay rate in narrow QWs with the growth of the barrier height.
As the height increases with growing of $x$,
the peak of $\hbar\Gamma_0$ becomes more pronounced and its maximum is shifted to the lower QW widths.
As we noted above, the peak at $L\approx 5$~nm for the GaAs/AlGaAs QWs is formed due to exciton squeezing by the QW potential. 
%
\begin{figure}[htbp!]%
\includegraphics*[width=\linewidth]{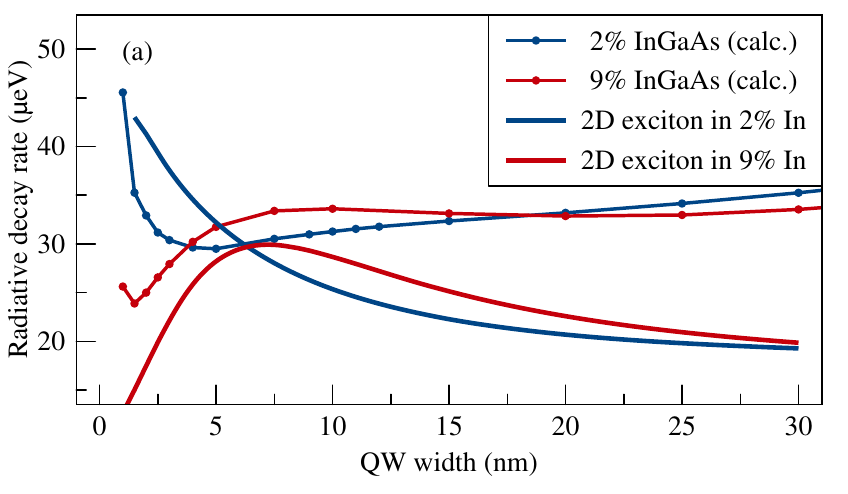}
\includegraphics*[width=\linewidth]{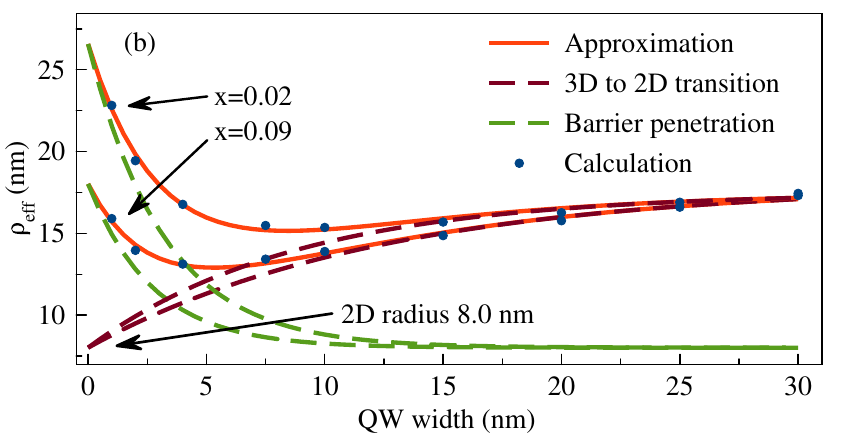}
\caption{%
\label{2drhoeffInGaAs}
(a) The calculated radiative decay rate the In$_x$Ga$_{1-x}$As/GaAs QWs with for 2\% and 9\% of indium (curves with points) in comparison with the 2D exciton model (smooth curves). Blue curves represent 2\% of indium in the heterostructures, while 9\% of indium are shown in carmine.
(b) The fitted effective 2D exciton radius $\rho_{\text{eff}}$ (solid points) for 2\%  and 9\% of indium as well as the phenomenological approximations (see the text).}
\end{figure}

In the case of In$_{x}$Ga$_{1-x}$As/GaAs QWs with small concentration of indium, $x = 0.02$, there is no peak of $\hbar\Gamma_0$ at the small QW widths. 
This is an indication of the weak exciton squeezing due to the small QW potential depth. 
%
For the In$_{x}$Ga$_{1-x}$As/GaAs QWs with  $x = 0.09$, some intermediate behavior is observed due to the intermediate hight of the barriers.
%
To understand this behavior, we applied the 2D exciton model described in Sect.~\ref{radiative_rate} and extracted $\rho_{\text{eff}}$ for  In$_x$Ga$_{1-x}$As/GaAs QWs with 2\% and 9\% of indium~[see Fig.~\ref{2drhoeffInGaAs}(b)].
The dependence $\rho_{\text{eff}}(L)$ for the shallow QW ($x = 0.02$) reveals the weak minimum, which is an indication of the weak squeezing of the exciton. Correspondingly, no maximum of $\hbar \Gamma_0$ should be observed for these QWs. Indeed, the calculated width dependence of $\hbar \Gamma_0$ shows monotonic rise with $L \to 0$, see Fig.~\ref{2drhoeffInGaAs}(a).
Curve $\rho_{\text{eff}}(L)$ for deeper QW ($x = 0.09$) shows the noticeable minimum at $L \approx 5$~nm that points out to an exciton squeezing.
As a result, a maximum of $\hbar \Gamma_0$ appears [Fig.~\ref{2drhoeffInGaAs}(a)].

The rapid growth of $\hbar \Gamma_0$ at $L \to 0$ in the InGaAs/GaAs QWs with 2\% of indium  [Fig.~\ref{2drhoeffInGaAs}(a)] is explained by the stronger penetration of exciton into the barriers as compared to the deeper InGaAs QWs with 9\% of indium. Due to the penetration, the overlap of the exciton wave function $\Phi(z)$ and the light wave [see Eq.~(\ref{Gamma0})] rises and the radiative decay rate increases as $L \to 0$.


In conclusion of this section, we compare our results of $\hbar\Gamma_{0}$ for InGaAs/GaAs QWs with those of Ref.~\cite{Andrea} for the same QW widths and indium concentration $x=0.09$.
The latter results have been also recently compared with experiment in Ref.~\cite{Poltavtsev2014}.
Our calculations show very similar dependence of the radiative decay rate in energy units, which, however, is shifted lower by 4--6~$\mu$eV.
Since the controlled precision of our calculations for $L < 50$~nm is better than 1.5~$\mu$eV, our results are improved in precision with respect to the earlier calculations.

\section{Discussion}
In the previous sections, we have shown that the exciton ground state as well as the radiative decay rate $\Gamma_{0}$ can be calculated by the direct numerical solution of the exciton SE.
The binding energy is obtained quite precisely by the direct numerical solution as well as by the variational approach with prescribed trial function~(\ref{v2}).
The good agreement
of the results of
two numerical methods
indicates that the chosen trial function is appropriate for the wide range of QW thicknesses.
It seems that, for each QW width, the varying parameters allow one to properly scale the trial function and, thus, to accurately simulate the exact wave function of the exciton ground state.
Therefore, besides the exciton binding energies, the wave functions obtained by two methods should also be very similar.

\begin{figure}[htbp!]%
\includegraphics*[width=\linewidth]{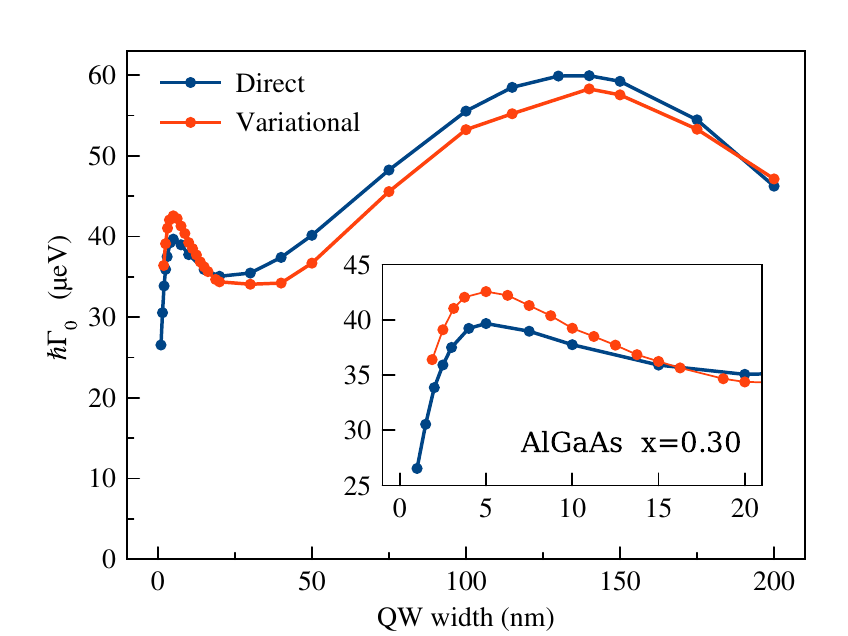}
\caption{%
\label{Discuss}
The radiative decay rate in energy units obtained for GaAs/AlGaAs QWs by the precise microscopic calculation in comparison to the results from the variational approach as a function of the QW width.}
\end{figure}

In this context, it is interesting to compare the radiative decay rates obtained using these two methods.
The exciton ground state wave function obtained by the variational approach for the GaAs/Al$_{0.3}$Ga$_{0.7}$As heterostructure has been used to calculate $\Gamma_{0}$
and comparison of $\hbar\Gamma_{0}$ is presented in Fig.~\ref{Discuss}.
One can see that the values of $\hbar\Gamma_{0}$  calculated by both methods differ slightly for the wide range of QW widths.
This difference of about 10\%, however, is
larger than the difference of the exciton binding energies at these QW widths.
In particular, for $L \sim 5$~nm the variational approach gives greater radiative decay rate than the direct numerical solution.
Nevertheless, such difference
is generally not so important for comparison with experimental data because the spread of our experimental data as well as of recent measurements (see Refs.~\cite{Poltavtsev,Poltavtsev2014}) substantially exceeds this discrepancy.
As a result, the variational approach with the proposed in Ref.~\cite{bib5} anzatz~(\ref{v2}) is appropriate for calculation of the exciton binding energy as well as the radiative decay rate.

\begin{figure}[htbp!]%
\includegraphics*[width=0.8\linewidth]{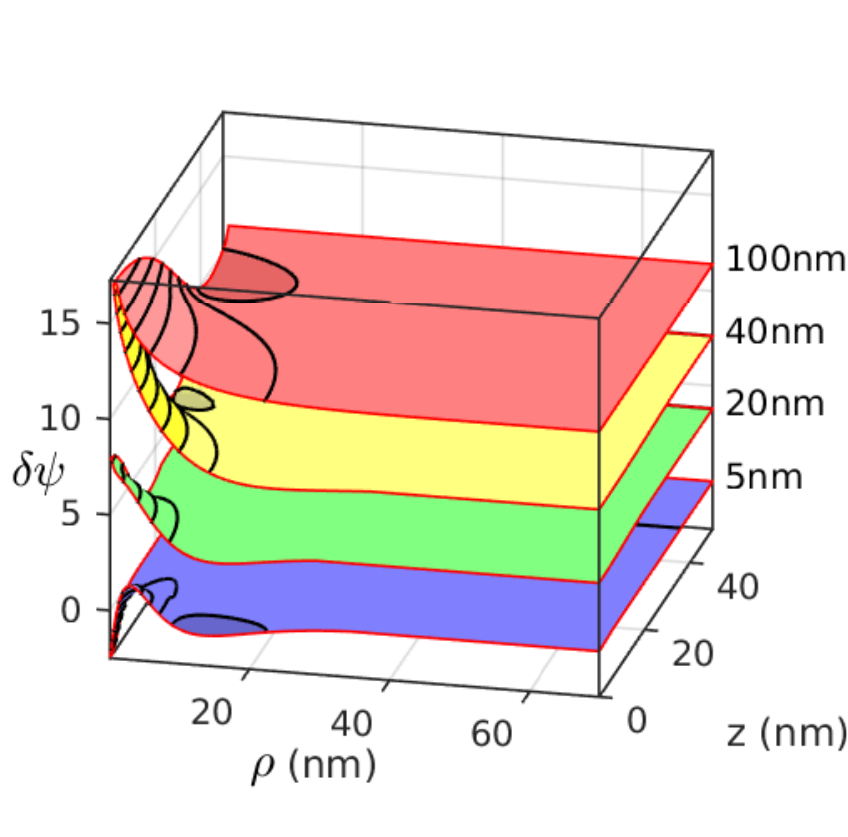}
\caption{%
\label{FigDiff}
The difference (in percents) of the wave functions of excitons in GaAs/AlGaAs QWs obtained by the precise microscopic calculation and by the variational approach as a function of two variables: $\rho$, $z=z_{e}=z_{h}$ for QW widths $L=5,20,40,100$~nm (from bottom to top). The functions are shifted vertically for visibility.}
\end{figure}

Until now, we have considered the exciton binding energy and radiative decay rate, which both are the integral characteristics of the exciton states.
Although the comparison showed us that these characteristics are similar, the wave functions obtained by different methods may differ in some details.
This difference may affect other properties of excitons, e.g., sensitivity to a magnetic field~\cite{Seisyan2012}.
So, for the future prospects we have compared the wave functions themselves.
Fig.~\ref{FigDiff} shows the difference (in percents) of the exciton wave functions obtained by two methods as a function of $\rho$ and $z$ for several widths of the GaAs/AlGaAs QWs.
It is clearly seen that the main difference of the wave functions is observed near the $z=0$, $\rho=0$ point.
This difference for small $L$ does not exceed 5\% and rapidly decreases with the $z$ and $\rho$ rise.
For large $L$, the difference is larger but also rapidly decreases as $z$ and $\rho$ grow.
%

\section{Conclusions}
In the present paper, we have numerically solved the three-dimensional SE and obtained the exciton binding energy as well as the ground state wave function. The studied SE was deduced from the electron-hole Hamiltonian with the Kohn-Luttinger term for the valence band taking into account only the heavy-hole excitons. The SE was solved by two methods: the direct one and the variational one. The variational method has been used by number of author who were studied this problem. The direct microscopic solution using the fourth-order finite-difference scheme has been carried out for the first time. It allowed us to precisely calculate the ground state and the exciton radiative decay rate.

We obtained the radiative decay rates for GaAs/AlGaAs and InGaAs/GaAs heterostructures for various QW widths $1 \le L \le 200$~nm and compound concentrations. We found that, since the different concentrations lead to the different magnitudes of band offsets, the behavior of the radiative decay rate for narrow QW widths ($2 < L \le 10$~nm) strongly depends on the concentration. Increase of the concentration leads to the higher barriers and to growth of the peak of the radiative decay. For very narrow QWs ($L \le 2$~nm) the radiative decay rate also behaves differently. If the QW is shallow, then the radiative decay rate grows as $L \to 0$. Instead, if the QW is deep, then it seems to be decreasing.

The obtained wave functions were used to test some simple models, namely the 3D exciton model (exciton in the bulk semiconductor) and the 2D exciton model. We have shown that these models are applicable only for wide and narrow QWs, respectively. The parameters of the models and the limits of applicability were estimated.

The comparison of the exciton binding energies obtained by the direct microscopic calculation and by the variational approach showed a very good agreement with the discrepancy of less than 0.1~meV for a wide range of QW widths. The analogous comparison of the radiative decay rates gave the similar overall behavior of these quantities, but larger differences (up to 10\%). Nevertheless, even for narrow QWs, this difference can be considered as insignificant because the available experimental data are more spread for these QWs. As a result, both numerical methods can be successfully used for calculation of the exciton binding energy and the radiative decay rate.

The experimental measurements of the radiative decay rate (HWHM) for several QW widths presented in the paper are consistent with the results of the precise calculations. The comparison with the earlier calculations showed slight systematic shift (of about $4-6$~$\mu$eV) of the earlier results with respect to our numerical data.

\begin{acknowledgments}
The authors would like to thank Alexander Levantovsky for providing the advanced plotting and curve fitting program MagicPlot Pro.
Financial support
from the Russian Ministry of Science and Education (contract no.
11.G34.31.0067), SPbU (grants No. 11.38.213.2014), RFBR and DFG in the frame
of Project ICRC TRR 160 is acknowledged. The authors also thank the SPbU
Resource Center ``Nanophotonics'' (www.photon.spbu.ru) for the samples
studied in present work.
Some calculations were carried out using the facilities of the ``Computational Center of SPbU''.
\end{acknowledgments}

\appendix*
\section{Numerical details}
In our study, we employ the well known finite-difference (FD) approximation~\cite{Collatz,Marchuk,Samarskii} of the derivatives in Eq.~(\ref{eq6})
due to its robustness as well as sparse and relatively simple form of the obtained matrix equation.
We consider the BVP for Eq.~(\ref{eq6}) at the domain $[-Z_{e}/2,Z_{e}/2]\times[-Z_{h}/2,Z_{h}/2]\times[0,R]$
over variables $z_{e}$, $z_{h}$, and $\rho$, respecitvely,
and specify the homogeneous boundary conditions at the boundaries.
The equidistant grids over each variable are introduced by the formulas
$z_{e,k}=k \Delta_{z_{e}}-Z_{e}/2$, $z_{h,l}=l \Delta_{z_{h}}-Z_{h}/2$, $\rho_{m} = m \Delta_{\rho}$,
where $\Delta_{z_{e,h}}=Z_{e,h}/(N_{z_{e,h}}+1)$, $\Delta_{\rho}=R/(N_{\rho}+1)$ are the grid steps and the indices $k$, $l$, $m$ go from 1 to some integer values $N_{z_{e}}$, $N_{z_{h}}$, and $N_{\rho}$, respectively.

\begin{widetext}
We use the central fourth-order FD formula for approximation of the second partial derivative of $\psi(z_{e},z_{h},\rho)$ with respect to $z_{e}$:
\begin{equation}
\label{fd0}
\frac{-\psi_{k-2,l,m}+16 \psi_{k-1,l,m}-30 \psi_{k,l,m}+ 16 \psi_{k+1,l,m} -\psi_{k+2,l,m}}{12\Delta^{2}_{z_{e}}}.
\end{equation}
Here, the unknown wave function on the grid $\psi(z_{e,k},z_{h,l},\rho_{m})$ is denoted as $\psi_{k,l,m}$.
The same FD formula is employed for the second derivative with respect to $z_{h}$.
The wave function $\psi$ at the knots beyond the considered domain over $z_{e}$ and $z_{h}$ is taken to be negligible due to its exponential decrease in this region.
We apply the noncentral fourth-order FD formulas for approximation of the first and second partial derivatives of $\psi(z_{e},z_{h},\rho)$ with respect to $\rho$:
\begin{equation}
\label{fd}
\begin{gathered}
\frac{-3\psi_{k,l,m}-10 \psi_{k,l,m+1}+18 \psi_{k,l,m+2}-6 \psi_{k,l,m+3}+ \psi_{k,l,m+4}}{12\Delta_{\rho}}, \\
\frac{10\psi_{k,l,m}-15 \psi_{k,l,m+1}-4 \psi_{k,l,m+2}+14 \psi_{k,l,m+3}-6 \psi_{k,l,m+4}+ \psi_{k,l,m+5}}{12\Delta^{2}_{\rho}}.
\end{gathered}
\end{equation}
\end{widetext}
It allows us to satisfy the trivial boundary condition at $\rho=0$ and to avoid knots $\rho_{m}<0$.
At the knots $\rho_{m}>R$, the wavefunction is also assumed to be zero.
It should be noted that this assumtion is possible only if the considered domain is large enough.

In the calculations, the grid steps over each variable have been taken to be the same, $\Delta=\Delta_{z_{e}}=\Delta_{z_{h}}=\Delta_{\rho}$, and multiply associated with the QW width.
The formulas~(\ref{fd0},\ref{fd}) define the theoretical uncertainty of the numerical solution of order of $\Delta^4$ as $\Delta\to 0$.
However, the discontinuity of the square potential~(\ref{eq2}) at the QW interfaces
decreases the convergence rate of the solution over $z_{e}$ and $z_{h}$ to order of $\Delta^{2}$,
whereas the convergence rate over variable $\rho$ is kept $\sim \Delta^4$.

\begin{figure}[htbp!]%
\includegraphics*[width=\linewidth]{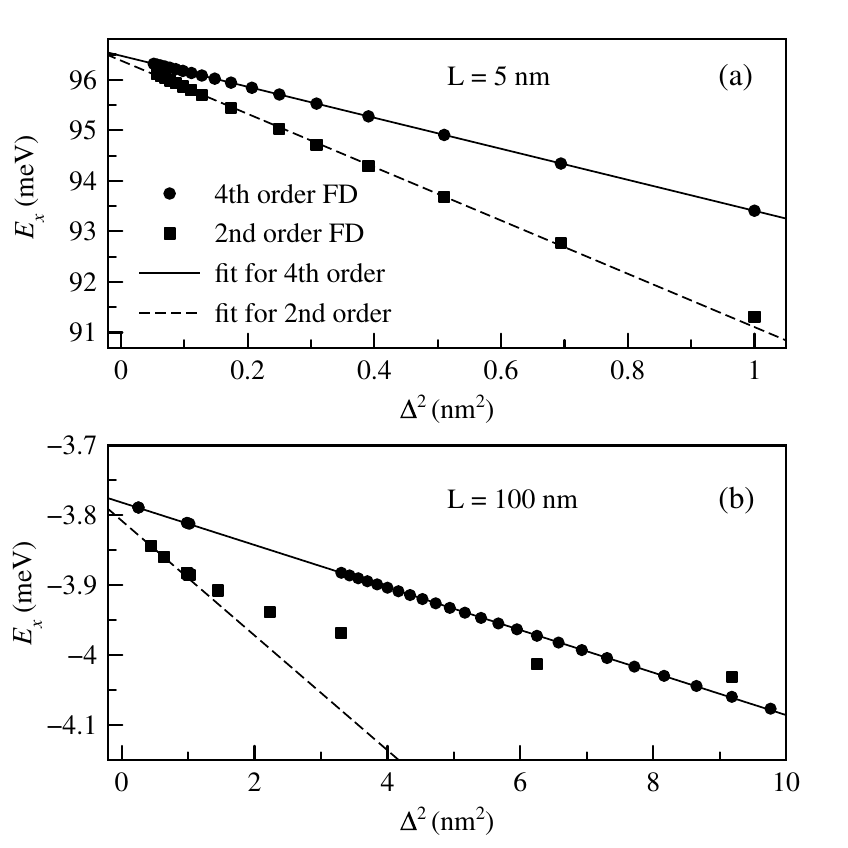}
\caption{The dependence of the exciton ground state energy, $E_{x}$, on the square of the grid step, $\Delta^{2}$, (a) for QW width $L=5$~nm and (b) for QW width $L=100$~nm.
The solid points are the calculated values of the energy.
The solid line indicates the fit of the energy calculated using the fourth-order FD scheme,
whereas the dashed line shows the possible extrapolation obtained using the second-order FD scheme.
}
\label{figh2h4lin}
\end{figure}

In order to choose the appropriate numerical scheme,
we compared the exciton ground state energy, $E_{x}$, calculated using
the second-order FDs and fourth-order FDs~(\ref{fd0},\ref{fd}) in Eq.~(\ref{eq6}).
The convergence of the exciton ground state energy as a function of the square of the grid step, $\Delta^{2}$, is presented in Fig.~\ref{figh2h4lin} for AlGaAs and QW widths $L=5,100$~nm.
Although both schemes provide convergence to almost the same values of energy as $\Delta\to 0$, the rate of convergence is different.
For the energy obtained using the second-order FD, the rate of convergence considerably depends on the width of QW.
For QWs of small widths (of order of the Bohr radius $a_{B}$), the linear convergence (with respect to $\Delta^{2}$) is observed for both schemes,
whereas for wide QWs the second-order scheme gives nonlinear convergence rate as $\Delta\to 0$.
The energy obtained using the fourth-order FD shows the linear dependence on $\Delta^{2}$
for the whole range of the studied QW widths. This linear behavior is not changed with increase of the QW width.
As a result, we performed a least square fit of the calculated energy and extrapolated it to $\Delta=0$.
The uncertainties estimated from the least square fit are quite small.
Thus, we obtained the precise exciton ground state energy.
The typical grid step reached in the calculations with the QW widths comparable with $a_{B}$ is $\Delta=0.25$~nm
whereas for wide QW the achieved grid step $\Delta = 0.5$~nm.

\begin{figure}[htbp!]%
\includegraphics*[width=\linewidth]{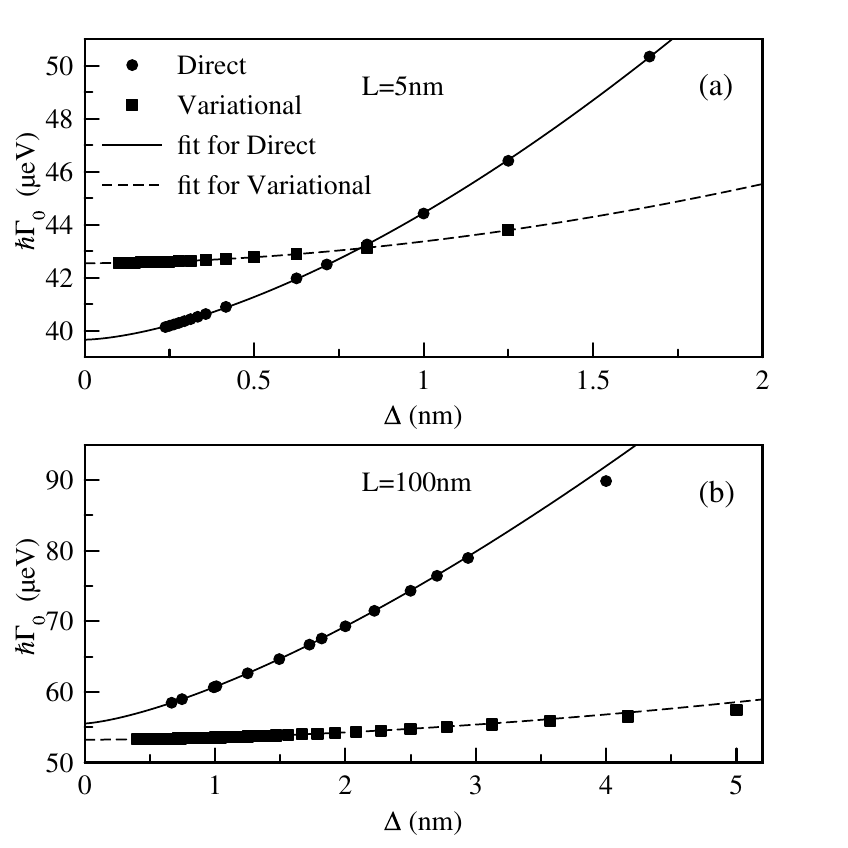}
\caption{The dependence of the radiative decay rate, $\hbar\Gamma_{0}$, on the grid step, $\Delta$, (a) for QW width $L=5$~nm and (b) for QW width $L=100$~nm.
The solid points are the calculated values of $\hbar\Gamma_{0}$.
The solid curves indicate the fit of the decay rate calculated using the direct numerical solution,
whereas the dashed curves show the fit of the values obtained using the variational approach.
}
\label{radconvergence}
\end{figure}

For the radiative decay rate in energy units, $\hbar\Gamma_{0}$, calculated using the ground state wave function obtained from the direct numerical solution,
the convergence is not so perfect.
The examples of the convergence rate for direct calculation of $\hbar\Gamma_{0}$ are shown in Fig.~\ref{radconvergence}.
For different QW widths, the convergence rate is different.
We were able to fit the convergence rate as $\Delta\to 0$ by the power function of $\Delta$ and estimate the uncertainty as the discrepancy of the calculated and extrapolated values.
This discrepancy was smaller for narrow QW than for the wide QW.
As a result, the uncertainty of our extrapolation for obtaining the precise radiative decay rate was estimated as $1.5$~$\mu$eV for QW width $L<50$~nm and $3$~$\mu$eV for wider QWs.
For the variational calculations, the convergence rate is good enough for precise extrapolation of the radiative decay rate.
This fact is also illustrated in Fig.~\ref{radconvergence}.

\end{document}